\documentclass[pmlr,twocolumn]{jmlr} 

\usepackage{booktabs}
\usepackage{diagbox}
\usepackage{enumitem}
\usepackage{subcaption}

\usepackage{multirow}

\title[Explainable Models for COVID-19]{Constructing and Evaluating an Explainable Model for COVID-19 Diagnosis from Chest X-rays}

\author{%
\Name{Rishab Khincha} \Email{f20160517@goa.bits-pilani.ac.in}\\
\addr BITS Pilani, Goa, India
\AND
\Name{Soundarya Krishnan} \Email{f20160472@goa.bits-pilani.ac.in}\\
\addr BITS Pilani, Goa, India
\AND
\Name{Tirtharaj Dash} \Email{tirtharaj@goa.bits-pilani.ac.in}\\
\addr BITS Pilani, Goa, India
\AND
\Name{Lovekesh Vig} \Email{lovekesh.vig@tcs.com}\\
\addr TCS Research, New Delhi, India
\AND
\Name{Ashwin Srinivasan} \Email{ashwin@goa.bits-pilani.ac.in}\\
\addr BITS Pilani, Goa, India
}

\begin{document}

\maketitle

\begin{abstract}
In this paper, 
our focus is on constructing
models to assist a clinician in the diagnosis of COVID-19 patients in situations where it is easier and cheaper
to obtain X-ray data than to obtain high-quality images like those from CT scans.
Deep neural networks have repeatedly been shown to be capable of constructing highly
predictive models for disease detection directly from image data.  However, their use in assisting
clinicians has repeatedly hit a stumbling block, because of their black-box nature. Some of this difficulty can be alleviated
if predictions were accompanied by explanations
expressed in clinically-relevant terms. In this paper, deep neural networks are used to extract domain-specific features(morphological features like ground glass opacity and disease indications like pneumonia) directly from the image data. 
Predictions about these features
are then used to construct a symbolic model (a decision tree) for the diagnosis of COVID-19 
from chest X-rays. We accompany these predictions with two kinds of explanations: visual (saliency maps,
derived from the neural stage),
 and textual (logical descriptions, derived from the symbolic stage). A radiologist rates the usefulness of the visual and textual explanations.
Our results suggest that:
(a) There is no significant loss in predictive accuracies when using the neural-symbolic model over using
a single end-to-end model; and
(b) The radiologist usually finds at least one of
    visual or textual explanations relevant;
	In addition, the
	textual explanations are better than
    a ``textual baseline'' more often than
    the corresponding visual case. This
    suggests
    either: the textual representation is
    is  more relevant to the diagnosistic task, or
    that the visual representation is not
    very relevant for the diagnostic task, or both.
Taken together, these results demonstrate
that neural models can be employed usefully
in identifying domain-specific features from
low-level image data; that textual
explanations in terms of clinically-relevant
features may be useful; and that visual
explanations will need to be clinically
meaningful to be useful.
\end{abstract}

\begin{figure*}[!htb]
\centering
\begin{minipage}{.43\textwidth}
    \centering
    \includegraphics[width=0.8\textwidth]{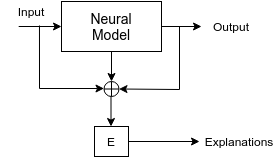}
    \subcaption{Single End-to-End Model}
    \label{fig:intro-bbox}
\end{minipage}%
\begin{minipage}{.53\textwidth}
    \centering
    \includegraphics[width=0.95\textwidth]{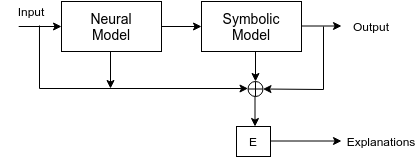}
    \subcaption{Hierarchical Model}
    \label{fig:intro-gbox}
\end{minipage}%
\caption{Two approaches to building explainable models}
\label{fig:intro-pipelines}
\end{figure*}

\section{Introduction}\label{sec:intro}
The use of explainable symbolic models for diagnosis
of disease is not new (MYCIN and Internist-I, for example, were developed from the
early 1970s, with just this purpose \cite{shortcliffe1975mycin,miller1982internist}).
Neither is the use of hierarchical approaches to
Computer Vision, in which high-level symbolic
models use the results of low-level feature-extraction
from raw images (see for example, \cite{ballard1982cv}).
For most of these, the symbolic component has
fulfilled purpose of generating limited answers to
``Why'' type  questions. Recent developments in
hardware, storage, and algorithmic optimisation has
however made possible the construction of very large
end-to-end neural network models that are capable of
predicting the probability of disease, directly from
raw image data (see Fig.~\ref{fig:intro-pipelines})

Although end-to-end models can
produce accurate diagnosis of COVID-19  \citep{khalid2020covid, mesut2020deep, mangal2020covidaid, wang2020covidnet, gozes2020rapid}, they are often considered as
black-box models, with little or no explanatory power
\citep{lipton2016interp}. This has proved to be a stumbling block
in their clinical acceptability. 





However, interpretability and explainability of these models has not been a focus for these works. Work in these areas is crucial for acceptance by physicians, as well as for sanity checks during model development.

Saliency-like visual explanations (GSInquire in \cite{wang2020covidnet} and GradCAMs in \cite{mangal2020covidaid, karim2020deepcovidexplainer}
have been used to highlight the relevant regions for the diagnosis made by the model. However, these maps may have less utility in a medical setting, consistent to previous findings like in \cite{arun2020assessing}. None of these works explore any other means of explanations. Most crucially, most works do not consult clinicians to find out which types of explanations are actually meaningful to them and assist them in diagnosis.



In this paper, we investigate
the use of a hierarchical neural-symbolic approach for
COVID-19 diagnosis, accompanied by a description
of why the diagnosis was reached. We consider
two kinds of explanations:
(a) Visual, in the form of saliency maps from
the neural-stage that construct the
domain-specific features used for diagnosis; and
(b) Textual, in the form of a description obtained
    from the symbolic model when making a diagnosis.
We compare these against the following baselines
that are agnostic to
any decision being made:
(c) A simple segmentation map of the lungs; and
(d) A simple tabulation of the input to the symbolic model for
    the data instance being diagnosed.
The thinking here is that for (a) to be a relevant visual
explanation, it should be judged (by a radiologist)
to be at least better than (c); and for (b) to be relevant,
it should be judged (again, by a radiologist) to be
at least better than (d).

Recently, a categorisation of model explainability has 
been proposed by \cite{arrieta2019explainable}.
In this categorisation,
explanations in (a) constitute local, visual explanations;
and (b) constitute global, textual explanations. To promote reproducibility of our findings, and for any future extensions, we provide the datasets collected from an experienced radiologist on request.


\begin{figure*}[!htb]
\centering
\includegraphics[width=0.9\textwidth]{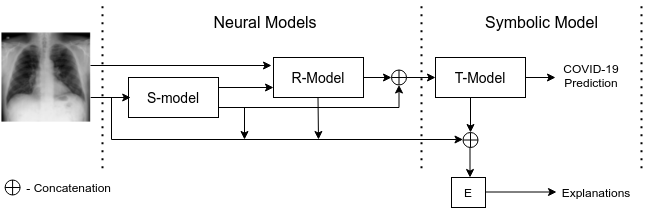}
\caption{Neural-Symbolic Pipeline for COVID-19 diagnosis. The
$S$-model is a deep network for predicting symptoms; the
$R$-model is a deep network for predicting radiological
features; and the $T$-model is a decision-tree. $E$ refers
to a set of procedures for extracting visual and symbolic
explanations.}
\label{fig:neuro-symbolic-pipeline}
\end{figure*}
\section{Neural-Symbolic Modelling for Diagnosis and Explanation}
\label{sec:neurosym}


In this section, we describe a hierarchical approach for
obtaining diagnoses and explanations.
The first stage of the hierarchical approach
performs a form of domain-specific representation-learning. Two
deep neural networks are used to obtain clinically relevant features directly from images.
The $S$-model is a deep network
that predicts symptomatic features like consolidations,
pleural effusions  and so on. The $R$-model is a deep network
that predicts morphological features like ground-glass opacity and
air-space opacifications (see Fig.~\ref{fig:neuro-symbolic-pipeline}). The principal datasets used in each of the models are
shown in Table~\ref{fig:data}.

\begin{figure*}[!htb]
\centering
\includegraphics[width=0.9\textwidth]{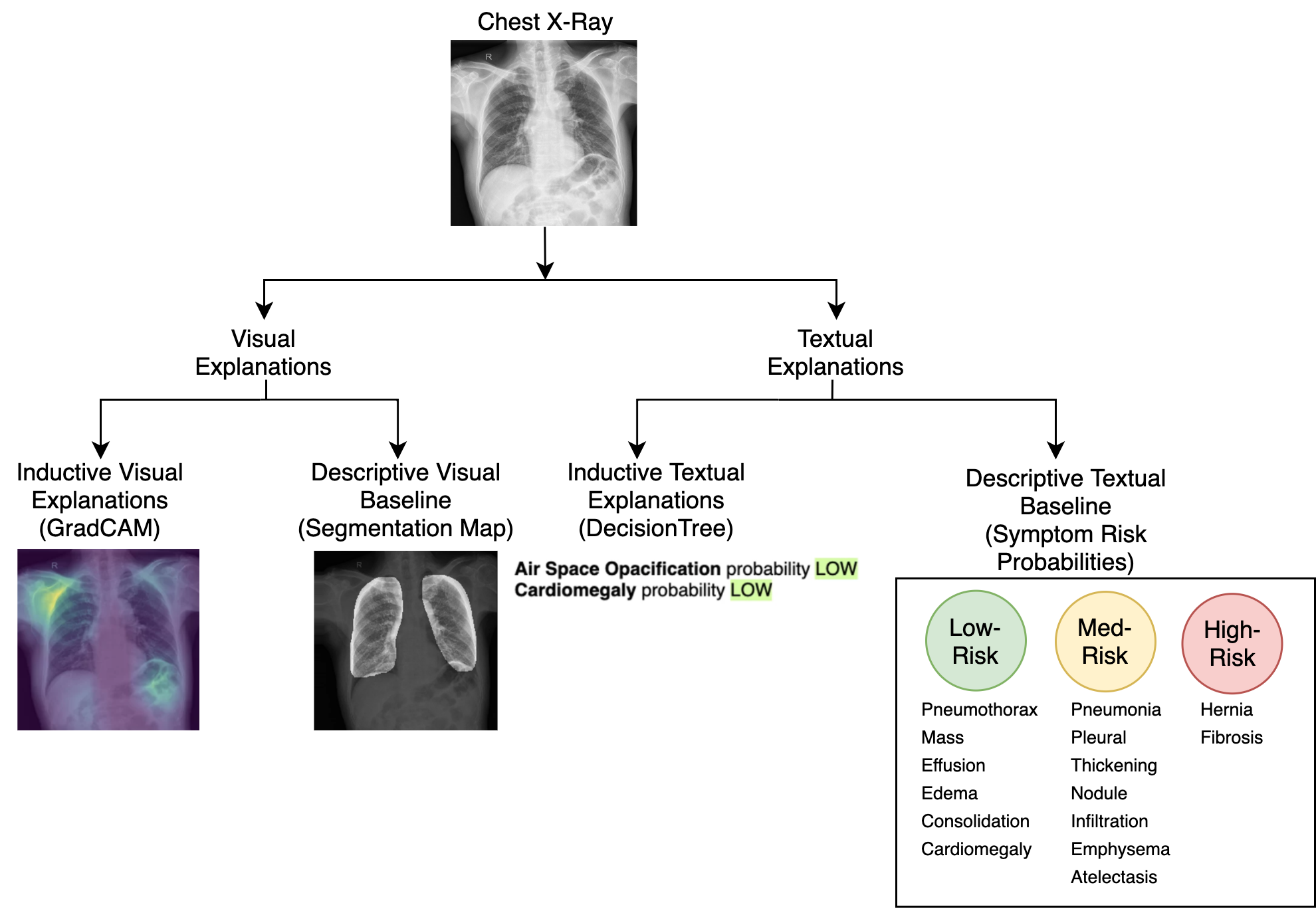}
\caption{Explanations for COVID-19 diagnosis}
\label{fig:explanations}
\end{figure*}

\subsection{Stage 1: Neural Models for Clinical Features}
\label{sec:neural}

The $S$-model is a CheXNet \citep{rajpurkar2017chexnet} model, obtained by
minimising the binary crossentropy loss using an Adam optimizer.

The $R$-model  is a deep model that identifies
radiologically-relevant morphological symptoms. The
model in this paper focuses on predicting the occurrence
 of various forms of opacification in chest X-rays.
To train the $R$-model, we use a dataset assembled
from: (a) Morphological annotations for COVID-19 X-rays
from the COVIDx dataset - we call this the COVIDr dataset; (b)
Images for pneumonia patients taken from the
NIH dataset; (c) Images for healthy and
tuberculosis patients taken from the
Pulmonary Chest X-ray dataset.
The COVIDr dataset contains annotatations made by a practising radiologist from the NHS, UK. We describe the annotation process first.

Chest X-rays from COVID-19 positive patients are
annotated with one of 7 morphological labels. Here,
we are interested in simply two broad categories
of these labels, namely the presence (or absence)
of Ground Glass Opacity ($GGO$) and Air Space Opacification ($ASO$). In COVID-19 images, $GGO$ is $present$
if the Ground Glass Opacity annotation is marked as
present; and $ASO$ is $present$ if any of:
Bilateral or Unilateral Air Space Opacification annotations
are marked as \emph{present} (See Appendix~\ref{apd:covidr} for more details on the COVIDr dataset). For healthy samples, we assume both $GGO$ and
$ASO$ are absent. For tuberculosis and pneumonia
patients, both $GGO$ and $ASO$ are not known (these images
were not seen during annotation).

The $R$-model is constructed to predict a $1$-hot encoding of
of one of 5 values: only $GGO$ is $present$; only $ASO$ is $present$;
both $GGO$ and $ASO$ are $present$; both $GGO$ and $ASO$ are $absent$; and both $GGO$ and $ASO$ are unknown. The model
is constructed using a training, validation, and the independent held-out test data as described
Section \ref{sec:method}. Input data for constructing
the model consist of
chest X-ray images, and the predictions of the
$S$-model. The
final model is obtained
by minimising the categorical cross-entropy loss using an Adam optimizer.

\begin{table*}[htb!]
\centering
\caption{Datasets Table}\label{fig:data}
\begin{tabular}{lll}
\toprule
Dataset & Model & Details \\ \midrule
\multirow{3}{*}{\begin{tabular}[c]{@{}l@{}}NIH Chest X-ray\\ \cite{wang2017nih}\end{tabular}} & $S$-model & \begin{tabular}[c]{@{}l@{}}112,120 X-ray images with 14 disease\\ labels from 30,805 unique patients\\ Default train-val-test split used.\end{tabular} \\ \cmidrule{2-3} 
 & $R$-model & \begin{tabular}[c]{@{}l@{}}Pneumonia images used. \\ Train(234) and test(88).\end{tabular} \\ \midrule
\multirow{4}{*}{\begin{tabular}[c]{@{}l@{}}Pulmonary Chest X-ray \\ \cite{jaeger2014pulmonary}\end{tabular}} & $Seg$-model & \begin{tabular}[c]{@{}l@{}}Chest X-rays and their masks used. \\ Train (566) and Test (134)\end{tabular} \\ \cmidrule{2-3} 
 & $R$-model & \begin{tabular}[c]{@{}l@{}}Tuberculosis and healthy Chest X-rays used. \\ Tuberculosis - Train (295) and Test (88); \\ Healthy - Train(294) and Test(110)\end{tabular} \\ \midrule
\multirow{3}{*}{\begin{tabular}[c]{@{}l@{}}COVIDx \\ \cite{cohen2020covid19}\end{tabular}} & $Seg$-model & \begin{tabular}[c]{@{}l@{}}Chest X-rays and their manually \\annotated masks used.\end{tabular} \\ \cmidrule{2-3} 
 & $R$-model & \begin{tabular}[c]{@{}l@{}}COVID-19 positive Chest X-rays used. \\ Train (196) and Test (30)\end{tabular} \\ \midrule
 COVIDr & $R$-model &  \begin{tabular}[c]{@{}l@{}}Relevant radiological annotations from a\\practicing radiologist for COVID-19 positive\\Chest X-rays from COVIDx.\\See Appendix~\ref{apd:covidr} for details.\end{tabular} \\ 
\bottomrule
\end{tabular}
\end{table*}

\subsection{Stage 2: Symbolic Model for Diagnosis}
\label{sec:sym}

In this stage, outputs of the $S$ and $R$-models
along with labels for images are used to construct
a decision-tree for COVID-19 diagnosis (the class-labels
for the decision-tree are \emph{COV+} or \emph{COV--}, where
the latter refers to images of healthy, pneumonia and
tuberculosis patients).\footnote{
The input to for tree-construction are
probability vectors for symptoms and
boolean-values for the morphological features.
Recall the $R$-models essentially predict
whether $GGO$ and $ASO$ are present, absent, or missing.}
Details of tree-construction are in Section \ref{sec:method}.

\subsection{Explanations}
\label{sec:expl}

We examine two kinds of model-based explanations
visual and textual, each with the corresponding baselines.
We use the term ``inductive'' to qualify
the explanation based on a machine-constructed model. The baselines are simple summarisations  of the relevant features of the data, and are qualified by the term ``descriptive''.
We require the inductive explanations to be at
least as useful to the clinicians as the descriptive
summaries. Details of these categories are:

\begin{description}
\item[Visual, Inductive.] This consists of class
    activation maps (GradCAMs, proposed by \cite{selvaraju2016gradcam}), showing gradations of
    saliency in the $R$-model.
\item[Visual, Descriptive.] This acts of a baseline
    for the corresponding inductive explanation.
    It consists of masks showing
    a segmentation of the lungs obtained using a deep
    network. 
    \item[Symbolic, Inductive.] This is a translation of
    the sequence of decisions made by the symbolic model
    in reaching a diagnosis.\footnote{
    For simplicity, we leave out statements about missing
    values when generating explanations. It is unclear
    whether statements of the form ``X is missing'' 
    are of value to the radiologist, but such an expression
    can nevertheless have predictive value.}
    \item[Symbolic, Descriptive.] This acts
        as a baseline for assessing the utility
        of the corresponding inductive explanation.
        It is a tabulation of discretised values
        of probabilities of the outputs of the $S$ and $R$-models. The discretisation is into
        3 simple categories ($high$, $medium$,
        and $low$).
\end{description}
Examples of each of these kinds of these categories are shown
in Fig.~\ref{fig:explanations}.

\section{Empirical Evaluation}
\label{sec:expt}
\subsection{Aims}
\label{sec:aims}

Our aim is to test the neural-symbolic system proposed in
Section \ref{sec:neurosym} along the following dimensions:
\begin{itemize}
\item \textbf{Predictive} We investigate if there is any significant
    loss in predictive accuracy over using a single
    end-to-end deep network model with access to the
    same image data.
\item \textbf{Explanatory} We investigate the clinical
    usefulness of the visual and symbolic explanations
    described earlier.
\end{itemize}


\subsection{Materials}
\label{sec:mat}

\subsubsection{Data}

The datasets used in this paper are
shown in Table~\ref{fig:data}. The COVIDr dataset has been annotated a practising radiologist. It captures important ground glass opacities and air space opacifications from the Chest X-rays relevant to the diagnosis of COVID-19 (See Appendix \ref{apd:covidr} for more details).


\subsubsection{Algorithms and Machines}


All the experiments are conducted in Python environment (Tensorflow/Keras) in a machine with 64GB main memory, 16-core Intel processor and 8GB NVIDIA P4000 graphics processor.

\subsection{Methods}
\label{sec:method}

We adopt the following method for
evaluation of predictive accuracy:

\begin{enumerate}
    \item Split the COVIDx dataset into \emph{COVIDxTrain}
        and \emph{COVIDxTest}
    \item Construct an end-to-end deep model on the \emph{COVIDxTrain}
        data and use it to obtain predictions on
        the \emph{COVIDxTest} data. \label{step:e2e}
    \item Construct an $S$-model using the $NIH$ dataset
        and use it to obtain predictions for
        the \emph{COVIDxTrain}  data. \label{step:smodel}
    \item Construct an $R$-model using pre-trained
        weights (see details below), and the
        \emph{COVIDxTrain} data and use it to construct
        predictions for morphological features
        for the \emph{COVIDxTrain} data. \label{step:rmodel}
    \item Construct a $T$-model (decision tree) using the outputs
        of the $S$-model (Step \ref{step:smodel})
        and the $R$-model (Step \ref{step:rmodel})
        and use it to obtain predictions on
        the \emph{COVIDxTest} data \label{step:tmodel}
    
    \item Compare the estimated accuracies on \emph{COVIDxTest}
        of the
        $T$-model (Step \ref{step:tmodel}) and
        the end-to-end deep model (Step \ref{step:e2e})
\end{enumerate}

\noindent
The following details are relevant:
\begin{itemize}
    \item We use the default split provided by \cite{cohen2020covid19}
        for splitting the COVIDx data in
        \emph{COVIDxTrain} and the completely held out test-set, \emph{COVIDxTest} 
    \item Pre-trained weights for the end-to-end
        model are from ImageNet. The learning rate used was $10^{-4}$ and the model was trained for 500 epochs.
    \item The train, validation and test set for the $S$-model uses the default split from \cite{wang2017nih} with 62428, 6336, and 1518 images respectively.
    \item There are two choices for pre-trained weights
        for the $R$-model. We have a choice of either
        using a generic pre-training (using ImageNet)
        or the domain-specific pre-training
        (using the end-to-end model in Step \ref{step:e2e}).
        We use a subset of \emph{COVIDxTrain} as a validation
        set to decide between these alternatives, which
        resulted in the use of domain-specific pre-trained
        weights.
    \item We split the \emph{COVIDxTrain} set into a 85-15 train-val       set to construct the $R$-model. We perform a sweep          across 3 learning rate settings ($10^{-3}$, $10^{-4}$ and      $10^{-5}$) and epochs (100, 250, 500) and the train split to build 9 $R$-models. The one performing best on the val     split is selected as our final $R$-model.
    \item Assessments of comparative difference in
        accuracies are made using a Gaussian approximation
        to the Binomial distribution. Given an
        accuracy of $p$ on \emph{COVIDxTest}, the
        standard deviation (s.d) is given by $\sqrt{p(1-p)/n}$,
        where $n$ is the number of instances in
        \emph{COVIDxTest}. Then a difference in accuracy
        is taken to be significant if it is
        2 or more s.d's away.
\end{itemize}

\noindent
We adopt the following method for evaluation
of explanations. For a subset of images in
\emph{COVIDxTest}:
 \begin{enumerate}
     \item Obtain inductive visual and textual explanations along with their corresponding baselines. 
            explanations (inductive and descriptive)
        \item Obtain a radiological assessment of
           (a) Utility  of visual and textual explanations;
           (b) Comparative utility of visual and textual
           explanations with the baselines.
\end{enumerate}

\noindent
The following details are relevant:

\begin{itemize}
    \item Visual
    inductive explanations are class
    activation maps obtained using
    GradCAM. Visual descriptive
    baselines are obtained using
    the lung segmentation model
    trained in Section~\ref{sec:expl}.
    \item Textual descriptive baselines are obtained from the predictions of the $R$ and the $S$-models by binning the probabilities into \emph{Low} ($\leq 0.33$), \emph{Medium} ($\leq 0.67$) or \emph{High}.
    \item The radiological assessments of the utility for each of the individual explanations as well as the comparative utilities are obtained by asking the radiologist a series of questions described in Appendix~\ref{apd:interface_expl}.
\end{itemize}

\subsection{Results}

\begin{table*}[!htb]
\centering
\caption{Confusion Matrices for COVID-19 Prediction. (COV$+$ is COVID-19 positive; COV$-$ is not COVID-19 positive)}
\label{fig:results-conf}
\begin{minipage}{.5\textwidth}
    \centering
    \subcaption{End to End model}
    \label{tab:results-bbox}
    \begin{tabular}{lcc}
    \toprule
    \diagbox{Act}{Pred} & COV$+$ & COV$-$ \\\midrule
    COV$+$ & 26 & 4 \\
    COV$-$ & 1 & 297 \\
    \bottomrule
    \end{tabular}
\end{minipage}%
\begin{minipage}{.5\textwidth}
    \centering
    \subcaption{Neural-Symbolic Model}
    \label{tab:results-ns}
    \begin{tabular}{lcc}
    \toprule
    \diagbox{Act}{Pred} & COV$+$ & COV$-$ \\\midrule
    COV$+$ & 23 & 7 \\
    COV$-$ & 2 & 296 \\
    \bottomrule
    \end{tabular}
\end{minipage}%
\end{table*}

\begin{table*}[htb!]
\centering
\caption{Usefulness of explanations and baselines as rated by the radiologist (Vis-Visual; Text-Textual; Ind-Inductive; Des-Descriptive)}
\label{tab:useful-all}
\begin{tabular}{lcccc}
\toprule
  & Vis-Ind & Vis-Des  & Text-Ind & Text-Des\\ \midrule
Useful & 14 & 21  & 17 & 6 \\
Somewhat Useful & 7 & 5 & 1 & 4\\
Not Useful & 9 & 1 & 12 & 20\\
\bottomrule
\end{tabular}
\end{table*}

The main observations to note are these:\footnote{Here, we
term an explanation as ``relevant'' if it is \emph{useful} or \emph{somewhat useful}.}
(1) The estimated predictive accuracy of
    the end-to-end approach is $0.985 \pm 0.007$
    and that of the Neural-Symbolic approach
    is $0.973 \pm 0.009$ (Table \ref{fig:results-conf});
(2) Visual explanations are marked as relevant
    on 21/30 instances. Of these, in
    5/21 instances the inductive explanation
    is marked as being more useful
    than the corresponding descriptive baseline (see Table \ref{tab:useful-quality-vis});
(3) Textual explanations are marked as relevant on
    18/30 instances. Of these, in 13/18 instances
    the inductive explanation is marked as being
    more useful than that the corresponding descriptive
    baseline (see Table \ref{tab:useful-quality-text});
(4) Either one of visual or textual explanation
    is marked as relevant in 29/30 instances. (See Table \ref{tab:useful-all} for details about the utility of each of the representations we provide. 


\begin{table*}[htb!]
\centering
\caption{Conditional comparison of inductive explanations ($I$) and descriptive baseline ($D$). ($X > Y$ means that $X$ is more useful than $Y$)}
\label{tab:useful-comp-VT}
\begin{minipage}{.48\textwidth}
\subcaption{Comparison given Visual Explanations (V) are relevant.}\label{tab:useful-quality-vis}
\centering
\begin{tabular}{lccc}
\toprule
 & Count  \\ 
\midrule
$I > D$ & 5  \\
$I < D$ & 8 \\
$I = D$ & 8  \\
\bottomrule
\end{tabular}
\end{minipage}
~
\begin{minipage}{.48\textwidth}
\subcaption{Comparison given Textual Explanations (T) are relevant.}\label{tab:useful-quality-text}
\centering
\begin{tabular}{lccc}
\toprule
 & Count  \\ 
\midrule 
$I > D$ & 13 \\
$I < D$ & 0  \\ 
$I = D$ & 5 \\
\bottomrule
\end{tabular}
\end{minipage}

\end{table*}



\begin{table*}[!htb]
\centering
\begin{minipage}{.48\textwidth}
    \centering
    \caption{Agreements between the model (M) and the radiologist (R) when the latter is sure}
    \label{tab:cliniciansure}
    \begin{tabular}{lcc}
    \toprule
    \diagbox[]{R}{M} & Agree & Disagree\\ \midrule
    Sure & 19 & 6 \\ 
    \bottomrule
    \end{tabular}
\end{minipage}%
~
\begin{minipage}{.48\textwidth}
    \centering
    \caption{Correct and Incorrect predictions of the model (M) when the radiologist (R) is unsure}
    \label{tab:clinicianunsure}
    \begin{tabular}{lcc}
    \toprule
    \diagbox[]{R}{M} & Correct & Incorrect \\ \midrule
    Unsure & 4 & 1 \\ 
    \bottomrule
    \end{tabular}
\end{minipage}%
\end{table*}



    \textbf{[Explanations are Useful.]} Either visual or textual
        explanations were  considered radiologically
        relevant in 29/30 instances. This
        suggests that it could be useful to
        augment clinical diagnoses with such
        additional information. \\
        
    \textbf{[Domain-Specific Representations are Important.]} There seems to be a reversal in preference from
    the descriptive baseline to the inductive explanation
    when going from visual to textual explanations.
    This is against a backdrop of a high frequency of
    preference for visual
    representations (21/30 instances). We believe several
    factors to be involved here. First, radiological
    diagnosis is largely image-based. This results in
    a natural preference for visual representations.
    However, the fact that the baseline of a simple
    lung segmentation is preferred over the saliency
    maps suggests that the latter do not correspond
    in any meaningful way. This is not the case
    with textual explanations that use
    radiologically meaningful features: hence the
    greater preference for the inductive explanation
    over the tabulation of descriptive statistics.
    Taken together, the results highlight the
    need to construct explanations in clinically
    meaningful terms. 
    This supports a finding by \cite{arun2020assessing} that
        direct presentation
        of GradCAM visualisation may not be `trustworthy', or relevant in a medical context. \\ \\
    \textbf{[``Simpler'' Explanations are Preferable.]}
        We think the apparent preference for visual baseline
        and textual inductive explanations 
        over the visual inductive explanation
        also indicate an underlying preference for
        simpler explanations. The visual
        baseline is simply a delineation of the lungs;
        and the preferred inductive textual explanation is a small
        set of conditions leading to the diagnosis. In
        contrast, the saliency maps and tabulations of
        probabilities contains significantly more detail,
        most of which is probably not thought to be
        clinically relevant. While
        the lung-segmentation image does not
        constitute any kind of explanation for the
        diagnosis of the model, 
        it is not entirely irrelevant,
        since radiologists do look at lung-appearance during
        diagnosis. \\ \\
\textbf{[Models Can Assist.]}
Our results in Tables \ref{tab:cliniciansure} and \ref{tab:clinicianunsure} suggest that the model is not
simply a clone of the radiologist, and could be helpful in cases where the diagnosis is not clear-cut. This opens up the possibility of using such systems as a second opinion when making diagnoses. \\

\section{Concluding Remarks}
In this paper, we have examined the use
of a hierarchical neural-symbolic approach
for diagnosis of COVID-19 from chest X-rays.
Our findings suggest that it is possible to construct
hierarchical models with accuracies comparable
to end-to-end models, with additional benefits
arising the ability to construct visual and textual
explanations using the neural and symbolic stages
of the approach. We have also presented some evidence
of an assessment of these explanations. Specifically,
we find that while explanations are relevant,
they are better than a simple ``baseline explanation''
only for the symbolic explanation that is a simple
logical description using domain-specific concepts.
This does not mean visual explanations are not meaningful:
rather that the specific use of saliency maps is not
useful.

A focus on explanations and keeping the clinician-in-the-loop is crucial for deployment of models as clinical assistants, so that clinicians gain trust in the decision of these machines, and possibly make better judgements. The deployment of quick and reliable model-based tools is becoming increasingly
important in the context of COVID-19, as it threatens to overwhelm medical practitioners in countries with
limited medical expertise.

It has been repeatedly shown in machine learning, that
accuracy of models depends crucially on the
representation used, and that the acceptance of 
machine-learnt models depends crucially on whether
the models are able to capture meaningful
mechanisms. Future work in this area needs to focus
therefore on the domain-specific features
to identify automatically from image-data, and how to
develop convey the concepts identified to the
radiologists in a manner in which they can
assess clinical significance. We
think this can only be achieved by
a continuous dialogue between clinicians and developers of machine-learning models.

\subsection*{Acknowledgements}
The authors would like to thank the radiologist from NHS, UK for helping us carry out this project.  
\clearpage
\bibliography{paper}

\clearpage
\appendix


\section{COVIDr Dataset}\label{apd:covidr}
\begin{figure*}[!htb]
\centering
\includegraphics[width=0.9\textwidth]{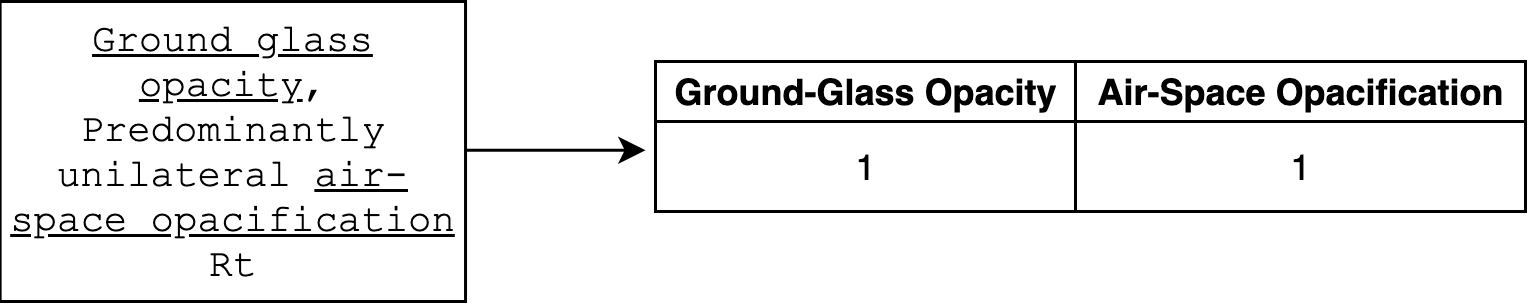}
\caption{Parsing the COVIDr dataset}
\label{fig:parse}
\end{figure*}

The details of the COVIDr dataset are seen in Table \ref{tab:covid-r}. An example of the way we parse the dataset is shown in Fig.~\ref{fig:parse}.

\begin{table*}[htb!]
\centering
\caption{The COVIDr dataset consists of 7 CXR finding labels for COVID-19 positive images. The number of positive instances of these labels are shown in the accompanying table. }
\label{tab:covid-r}
\begin{tabular}{ll}
\toprule
\textbf{CXR-Finding} & \textbf{Positive Instances} \\
\midrule
None & 44/245 (17.95\%) \\
Ground glass opacity & 145/245 (59.18\%) \\
Bilateral patchy air-space opacification & 30/245 (12.24\%) \\
Bilateral symmetrical air-space opacification & 18/245 (7.34\%) \\
Bilateral peripheral air-space opacification & 54/245 (22.04\%) \\
Predominantly unilateral air-space opacification Rt & 36/245 (14.69\%) \\
Predominantly unilateral air-space opacification Lt & 28/245 (11.4\%) \\ \bottomrule
\end{tabular}
\end{table*}






\section{Model Details}\label{apd:first}
\begin{figure*}[!htb]
\centering
\begin{minipage}{.53\textwidth}
    \centering
    \includegraphics[width=0.95\textwidth]{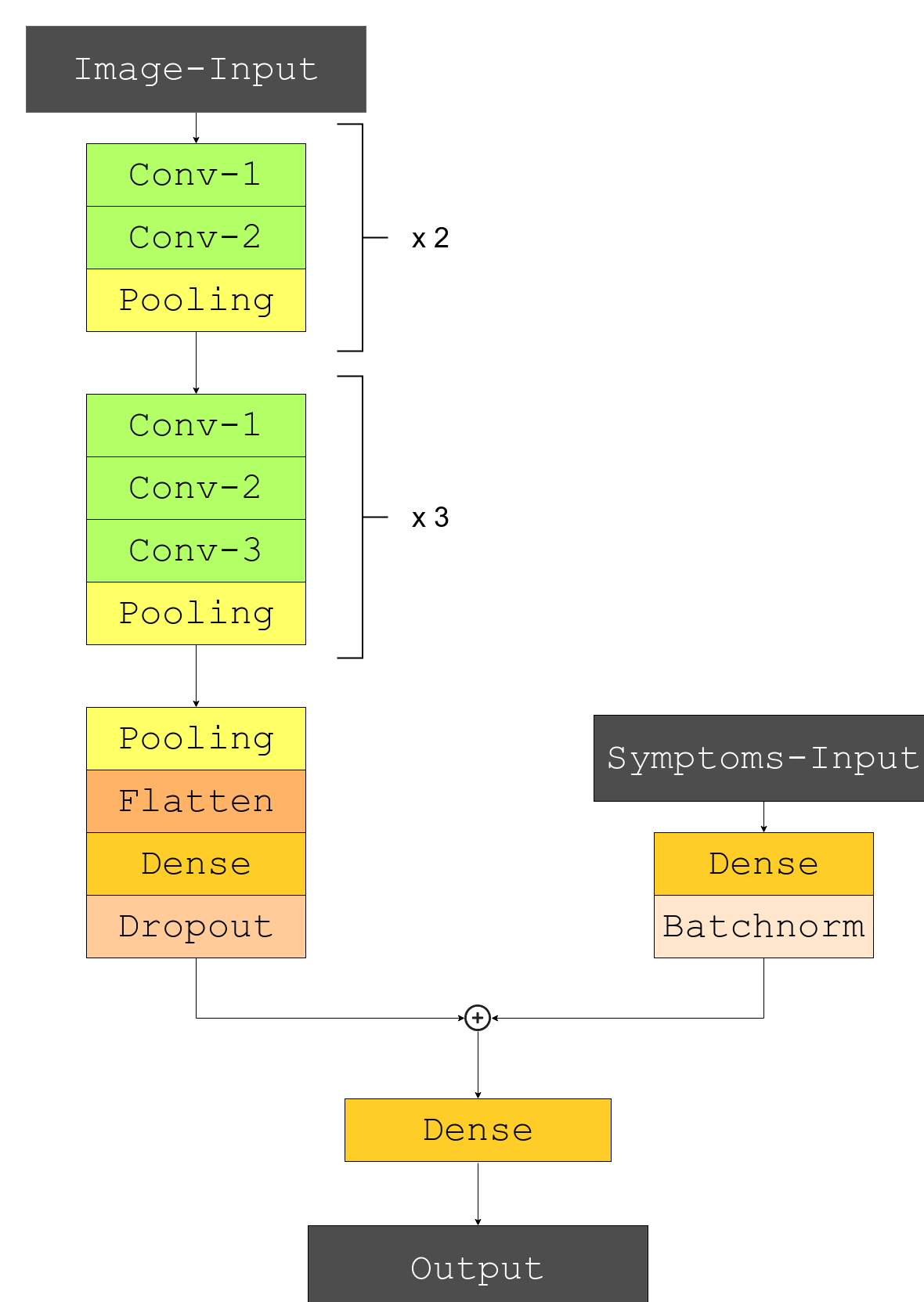}
    \subcaption{R-Model}
    \label{fig:RModel}
\end{minipage}%
\begin{minipage}{.53\textwidth}
    \centering
    \includegraphics[width=0.95\textwidth]{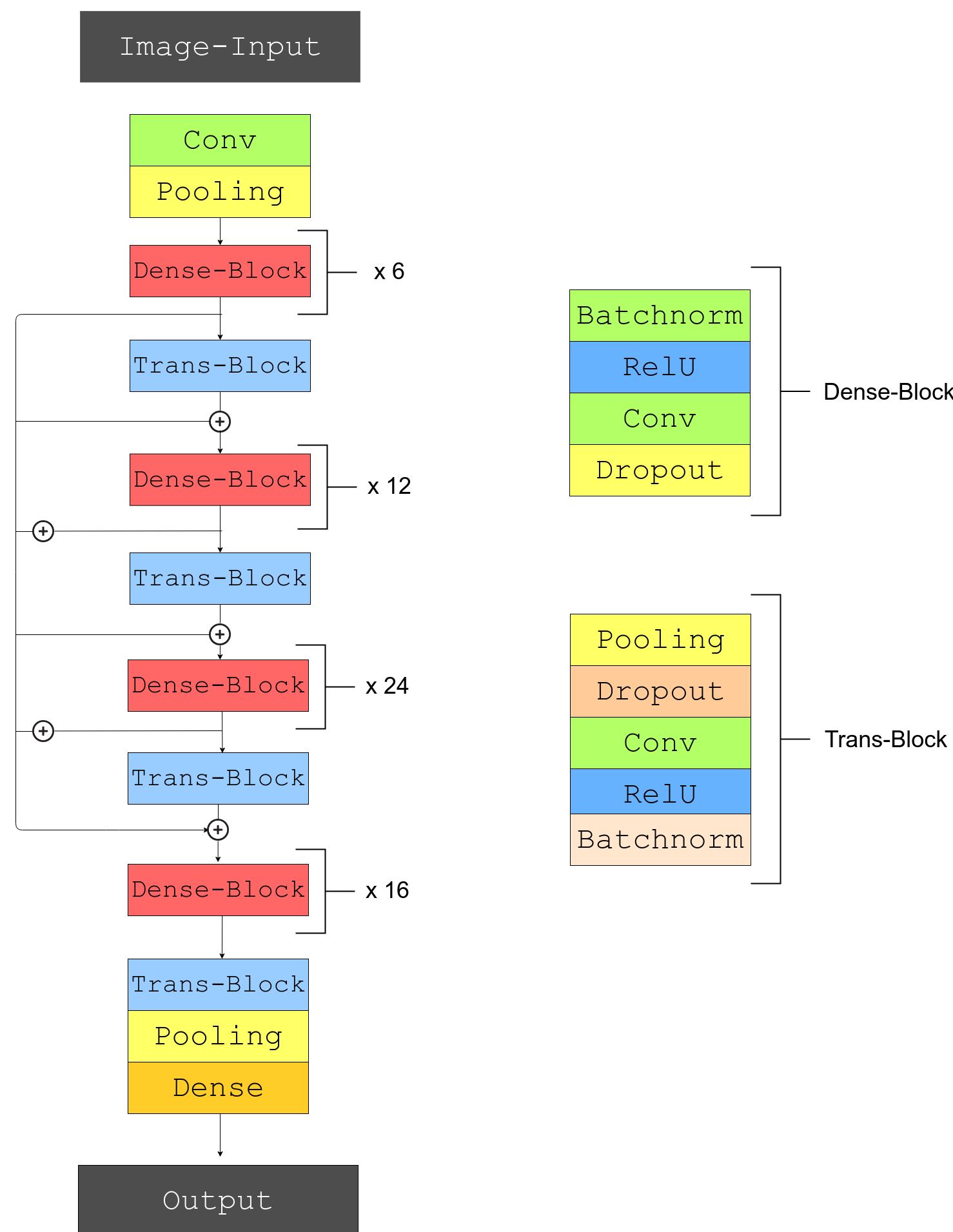}
    \subcaption{S-Model}
    \label{fig:SModel}
\end{minipage}%
\caption{Neural model architectures}
\label{fig:NeuralModels}
\end{figure*}

\begin{table}[htb!]
\centering
\caption{Conversion of Neural Radiology Model output to Symbolic Tree input}
\label{tab:ohe_to_mhe}
\begin{tabular}{lcc}
\toprule
Max Prob. & ASO & GGO \\ 
\midrule
ASO & 1 & 0 \\ 
GGO & 0 & 1 \\ 
ASO\_GGO & 1 & 1 \\ 
No\_ASO\_GGO & 0 & 0 \\ 
Missing\_ASO\_GGO & ? & ? \\ 
\bottomrule
\end{tabular}
\end{table}
The \emph{R-} and \emph{S-Model} architectures are shown in Fig.~\ref{fig:NeuralModels}. 
Details of conversion of the Neural Radiology Model ($R$-model) output to Symbolic Tree ($T$-model) input are shown in Table~\ref{tab:ohe_to_mhe}.

\begin{figure*}[!htb]
\centering
\begin{minipage}{.53\textwidth}
    \centering
    \includegraphics[width=0.9\textwidth]{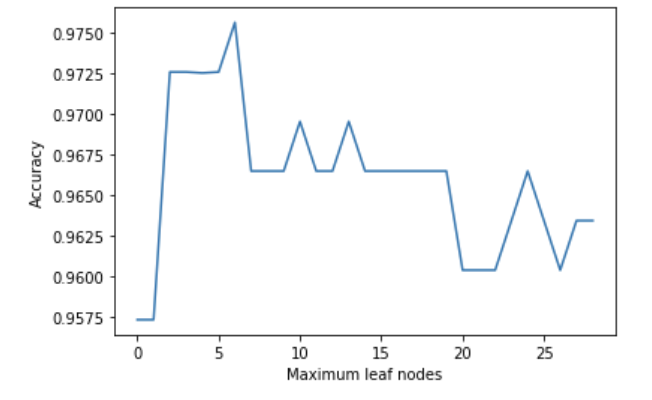}
    \subcaption{Accuracy vs. Max. Leaf Nodes}
    \label{fig:accvsleaves}
\end{minipage}%
\begin{minipage}{.53\textwidth}
    \centering
    \includegraphics[width=0.9\textwidth]{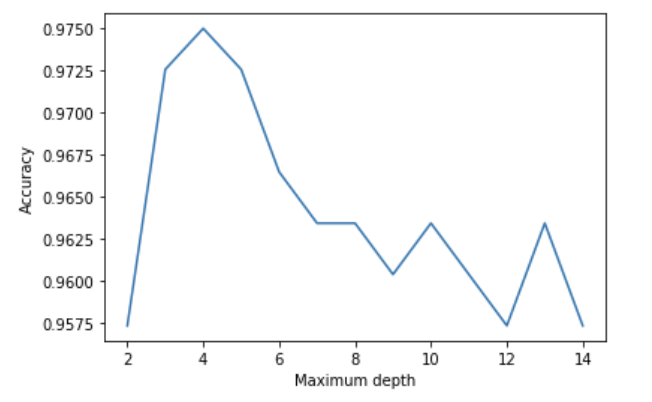}
    \subcaption{Accuracy vs. Max. Depth}
    \label{fig:accvsdepth}
\end{minipage}%
\caption{Optimal Leaf Nodes and Depth}
\label{fig:nspipeline}
\end{figure*}

To check for the optimal maximum leaf nodes, we plot (Fig.~\ref{fig:accvsleaves}) the test accuracy against the number of leaf nodes. We also try to find the optimal maximum depth of the tree, as shown in Fig.~\ref{fig:accvsdepth}. 


\begin{figure*}[!htb]
\centering
\includegraphics[width=0.9\textwidth]{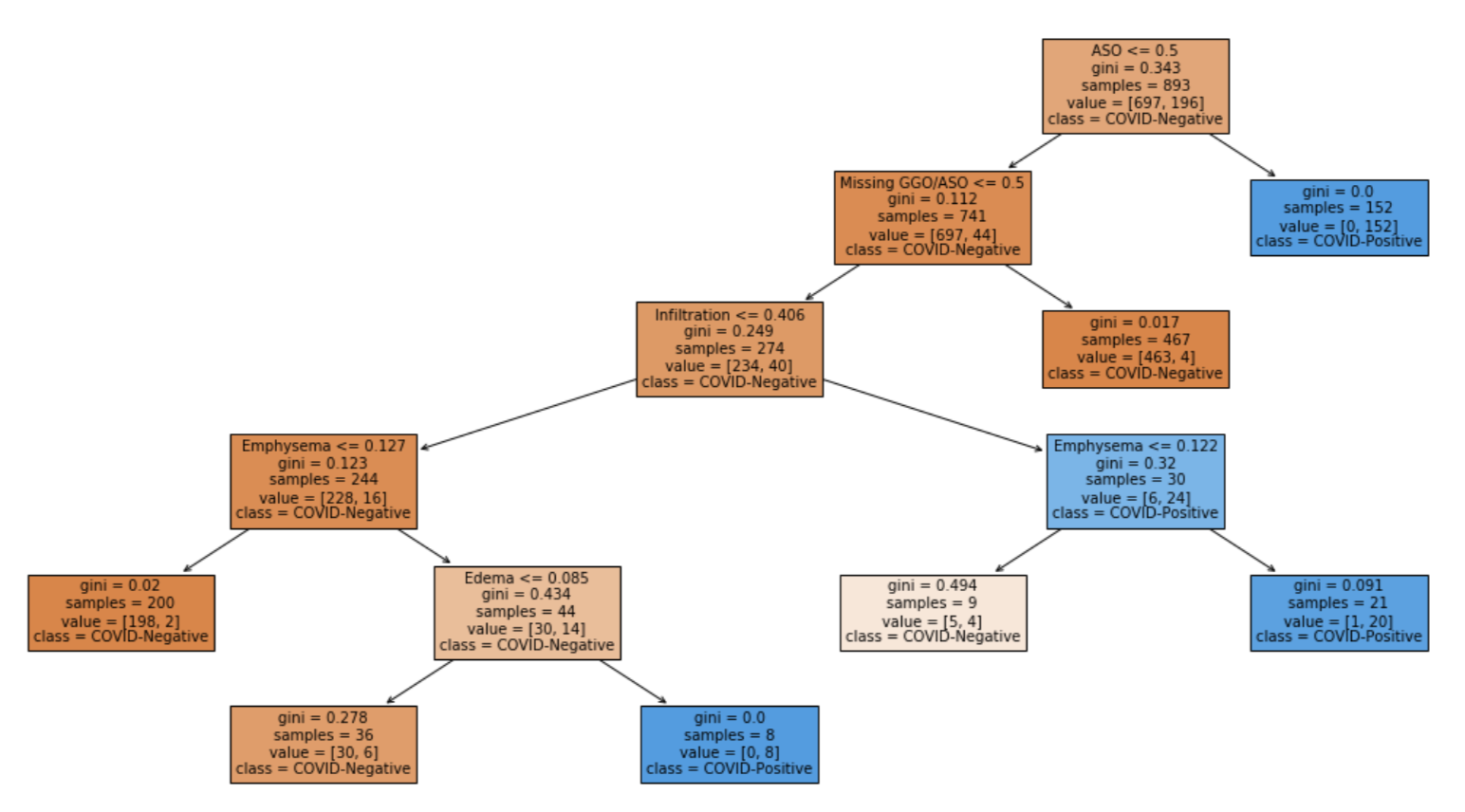}
\caption{Tree used to generate textual inductive explanations}
\label{fig:finaltree}
\end{figure*}

The tree obtained from the Symbolic model that is
described in Section~\ref{sec:intro} is shown in Fig.~\ref{fig:finaltree}. The extracted rules are 
shown below.\\

\noindent\textbf{Rules for COVID-Positive}

\begin{enumerate}
    \item \texttt{P(ASO) > 0.5}
    \item \texttt{P(ASO) <= 0.5 
    \\ \&\& P(Missing GGO/ASO) <= 0.5 
    \\ \&\& P(Infiltration) > 0.406 
    \\ \&\& P(Emphysema) <= 0.122}
    \item \texttt{P(ASO) <= 0.5
    \\ \&\& P(Missing GGO/ASO) <= 0.5 
    \\ \&\& P(Infiltration) <= 0.406 
    \\ \&\& P(Emphysema) > 0.127 
    \\ \&\& P(Edema) > 0.085}
\end{enumerate}

\noindent\textbf{Rules for COVID-Negative}
\begin{enumerate}
    \item \texttt{P(ASO) <= 0.5 
    \\ \&\& P(Missing GGO/ASO) <= 0.5}
    \item \texttt{P(ASO) <= 0.5  
    \\ \&\& P(Missing GGO/ASO) <= 0.5 
    \\ \&\& P(Infiltration) > 0.406 
    \\ \&\& P(Emphysema)>0.122 
    \\ \&\& P(Emphysema) <= 0.122}
    \item \texttt{P(ASO) <= 0.5 
    \\ \&\& P(Missing GGO/ASO) <= 0.5 
    \\ \&\& P(Infiltration) <= 0.406 
    \\ \&\& P(Emphysema)>0.127 
    \\ \&\& P(Edema) <= 0.085}
    \item \texttt{P(ASO) <= 0.5 
    \\ \&\& P(Missing GGO/ASO) <= 0.5 
    \\ \&\& P(Infiltration) <= 0.406 
    \\ \&\& P(Emphysema) <= 0.127}
\end{enumerate}   

\begin{figure*}[!ht]
\centering
\includegraphics[width=0.73\textwidth]{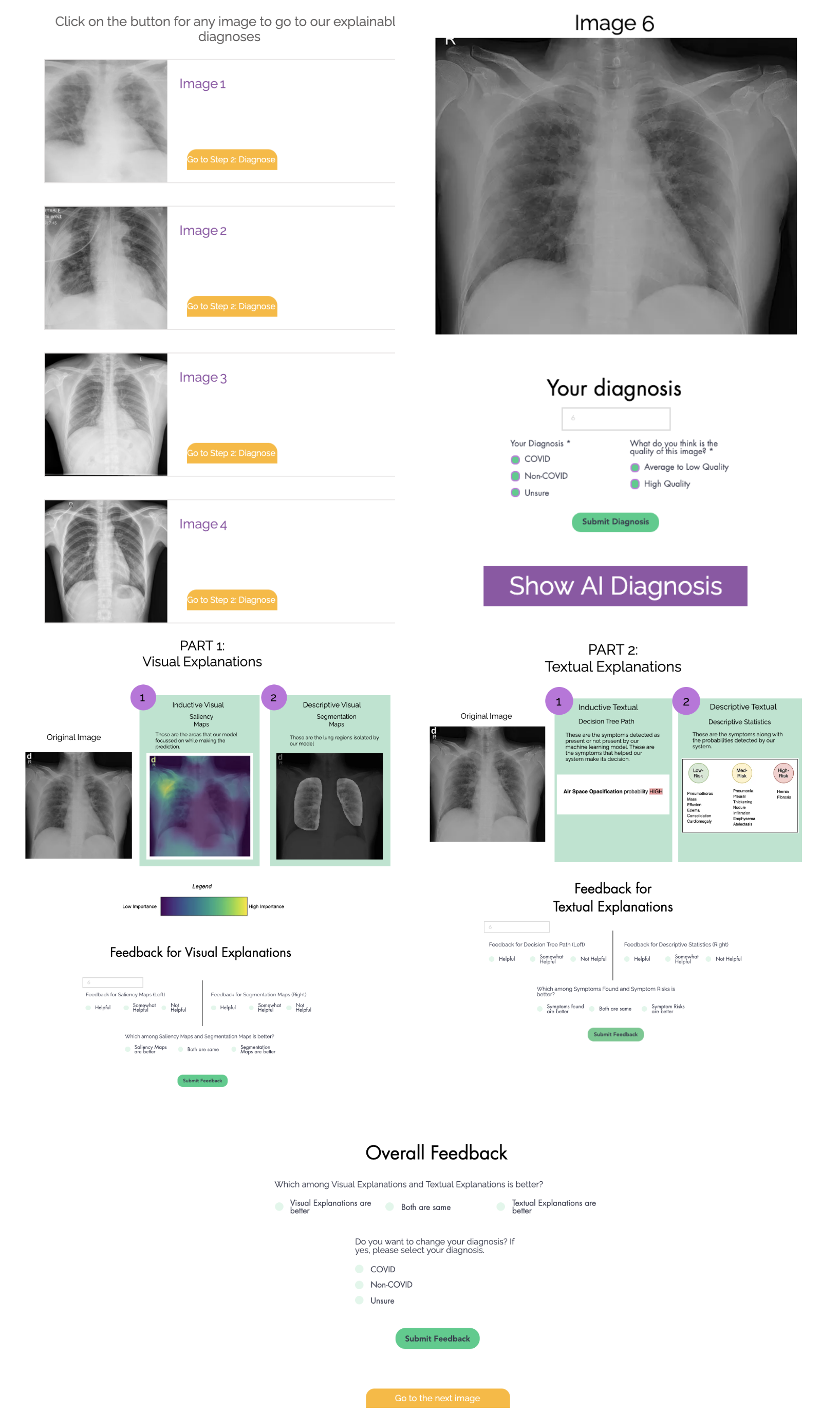}
\caption{Snippet of Interface that the Radiologist uses to evaluate explanations and baselines}
\label{fig:interface-overall-expla}
\end{figure*}


\section{Web-Based interface for collecting feedback}\label{apd:interface_expl}
A practising radiologist from the NHS, UK provides utility assessments for 30 Chest X-ray images sampled from the held-out test set (COVID-19 positive - 5, healthy - 5, tuberculosis - 5 and pneumonia - 5) , along with the COVID-19 diagnosis made by our model and all the explanations as stated in Section \ref{sec:expl}. We provide a web-based interface and ask the following questions in the given order :
\begin{enumerate}
    \item Prior to revealing the diagnosis of a given Chest X-ray image made by the model (true label is never revealed) or the corresponding explanations, we ask for the radiologist's diagnosis. We then reveal the diagnosis made by the model.
    \item We ask the radiologist to rate the visual quality of the image as Low, Medium or High. 
    \item The visual explanations(\emph{VisInd}, \emph{VisDed}) provided by our model are then displayed side by side along with the Chest X-ray. We then ask if each of these explanations individually are \emph{Useful},  \emph{Somewhat Useful} or \emph{Not Useful}. We also ask whether the \emph{VisInd} explanations was more useful, \emph{VisDed} explanations was more useful or if both are the same.
    
    \item The symbolic explanations(\emph{SymInd}, \emph{SymDed}) provided by our model are displayed side by side along with the Chest X-ray. We then ask if each of these explanations individually are \emph{Useful}, \emph{Somewhat Useful} or \emph{Not Useful}. We also ask whether the \emph{SymInd} explanations was more useful, \emph{SymDed} explanations was more useful or if both are the same. 

    \item Finally, we ask which explanations were more useful - Visual, Symbolic or both were the same.
\end{enumerate}
Snippets of the web interface are shown in Fig. \ref{fig:interface-overall-expla}


\section{Additional Results}\label{apd:addresults}
\subsection{COVIDr Baselines}
We benchmark the COVIDr dataset by training different variation to predict the morphological symptoms. The results are shown in Table \ref{tab:rmodel_res}.



\begin{table}[hbtp]
\centering
\caption{Results on the Radiology Model. (Sym-- Model using embeddings from the $S$-model. PT-- VGG16 uses pre-trained weights from the end-to-end model)}\label{tab:rmodel_res}
\begin{tabular}{lccc}
\toprule
Model & Accuracy & AUC \\ 
\midrule
VGG-16 & 0.83 & 0.96 \\
VGG-16 + Sym & 0.86 & 0.97 \\
VGG-16 + Sym (PT) & \textbf{0.88} & \textbf{0.99} \\ 

\bottomrule
\end{tabular}
\end{table}

\subsection{Symbolic Model Baselines}
The results of our symbolic model are shown in Table \ref{tab:tree_covid}. We compare these results with a single end-to-end, as well as other tree-based models. We find that the loss in accuracy in our symbolic model is not significant in comparison with the black box model, and the tree model additionally provides interpretable explanations. We also see that our decision tree has very similar predictive power as with other tree ensembling methods like \emph{Random Forest} and \emph{XGBOOST}.

\begin{table}[hbtp]
\centering
\caption{Results on the COVID-19 prediction}\label{tab:tree_covid}
\begin{tabular}{lccc}
\toprule
Model & Accuracy & F1 \\ \midrule
End-to-End & $\textbf{0.985} \pm \textbf{0.007}$ & \textbf{0.95} \\
NS-RF & $0.973 \pm 0.009$ & 0.915 \\
NS-XGB & $0.973 \pm 0.009$  & 0.930 \\
NS-DTree & $0.973 \pm 0.009$  & 0.915 \\
\bottomrule
\end{tabular}
\end{table}

\end{document}